\documentclass[PRD,amsmath,amssymb,preprintnumbers,nofootinbib,floatfix,uperscriptaddress]{revtex4-2}
\usepackage{graphicx}
\usepackage{bm}
\usepackage{hyperref}

\begin{document}

\title{Comment on arXiv:2608.29450: ``First Measurement of Solar Neutrinos through Elastic Neutrino-Electron Scattering at the keV Scale''}

\author{Ke Han$^{2,1}$}
\thanks{ke.han@sjtu.edu.cn}
\author{Xiangdong Ji$^{1,2}$}
\thanks{xdji@sjtu.edu.cn}
\author{Jianglai Liu$^{1,2}$}
\thanks{jianglai.liu@sjtu.edu.cn, spokesperson of the PandaX collaboration}

\collaboration{on behalf of the PandaX Collaboration}

\affiliation{\vspace{0.1cm}\\ 
$^1$Tsung-Dao Lee Institute, Shanghai, China, 201210\\
$^2$School of Physics and Astronomy, Shanghai Jiao Tong University, Shanghai, China, 200240}

\date{\today}

\maketitle

The XENONnT Collaboration recently reported a new measurement of the solar \(pp\) neutrino flux via neutrino-electron elastic scattering using a multi-ton-scale liquid xenon detector (preprint submitted on Aug. 29, 2026~\cite{xenon} and release talk on Aug. 31, 2026~\cite{xenon-seminar}). While we congratulate them on this definitive result, we comment here on an important omission in their summary plot (Fig.~6).

The PandaX Collaboration previously presented a preliminary analysis of the solar $pp$ neutrino flux at the Neutrino 2026 Conference (June 24, 2026, UC Irvine), with the slides available online~\cite{han_talk}. A subsequent preprint detailing the full analysis was published on arXiv on July 2, 2026 (arXiv:2607.02405)~\cite{pandax2026}. This updated result represents a significant improvement over our previous measurement published in 2024~\cite{pandax2024}.

While the XENONnT manuscript includes a ``Note Added'' acknowledging our recent preprint~\cite{pandax2026}, the summary plot (Fig.~6, top panel) omits this updated data point. Instead, it displays our older, superseded 2024 measurement~\cite{pandax2024}. 

\begin{figure}[htbp]
    \centering
    \includegraphics[width=0.60\textwidth]{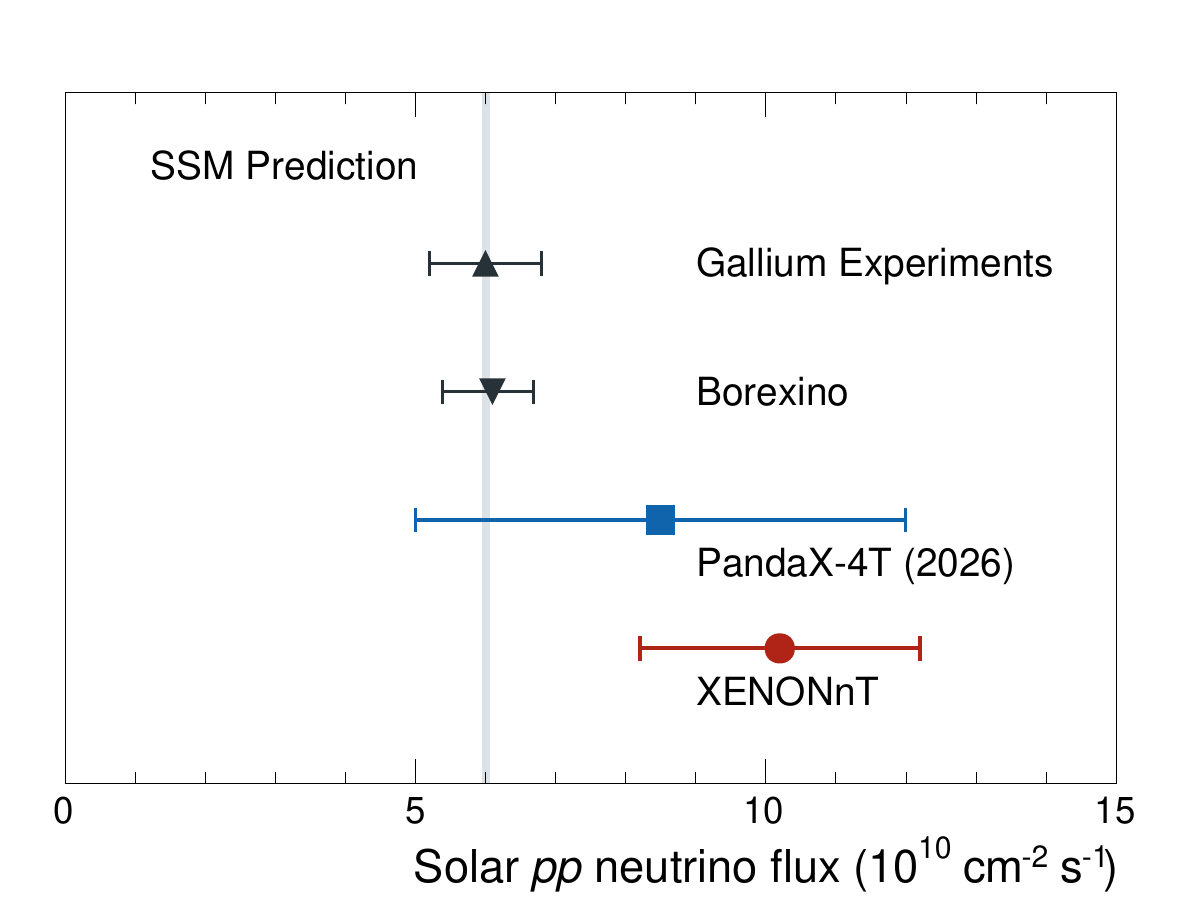}
    \caption{Comparison of solar $pp$ neutrino flux measurements from Gallium experiments~\cite{SAGE:2009eeu}, Borexino~\cite{BOREXINO:2018ohr}, PandaX-4T (2026)~\cite{pandax2026}, and XENONnT~\cite{xenon} against the Standard Solar Model (SSM) prediction based on high-metallicity (GS98)~\cite{Grevesse:1998bj} and low-metallicity (AGSS09)~\cite{Asplund:2009fu} models (vertical gray line).
    }
    \label{fig:comparison}
\end{figure}

To provide a complete and accurate record, an updated comparison is presented in Fig.~\ref{fig:comparison}. This plot explicitly displays the PandaX-4T (2026) result alongside the XENONnT, Borexino, and Gallium data against the Standard Solar Model (SSM) prediction. 

\end{document}